\documentstyle[aps,epsf]{revtex}
\begin{document}
\title{
Simulation of Lattice Polymers with Multi-Self-Overlap Ensemble
\footnote{submitted to J. Phys. Soc. Jpn.}
}
\author{Yukito Iba\footnote{e-mail:iba@ism.ac.jp}}
\address{
Department of Prediction and Control,
       The Institute of Statistical Mathematics,\\
       4-6-7, Minami-Azabu, Minatoku 106-8569, Tokyo, Japan  
}
\author{George Chikenji\footnote{e-mail:chikenji@hyperion.phys.sci.osaka-u.ac.jp}
, and Macoto Kikuchi}
\address{
      Department of Physics, Osaka University, Toyonaka 560-0043, Japan
}
\maketitle
\vspace*{2cm}
\begin{abstract}
A novel family of dynamical 
Monte Carlo algorithms for lattice polymers is proposed. 
Our central idea is to simulate
an extended ensemble in which 
the self-avoiding condition
is systematically weakened. 
The degree of the self-overlap is
controlled in a similar manner as the multicanonical ensemble.
As a consequence, the ensemble
---the multi-self-overlap ensemble---
contains adequate portions
of self-overlapping conformations as well as higher  
energy ones. 
It is shown that the multi-self-overlap ensemble algorithm reproduce
correctly the canonical averages at finite temperatures
of the HP model of lattice proteins.
Moreover, it outperforms massively a
standard multicanonical algorithm
for a difficult example of a polymer with 8-stickers.
Alternative algorithm based on exchange Monte Carlo method
is also discussed.
\end{abstract}

\twocolumn
 
Monte Carlo simulations of lattice polymers has been an important subject
in a wide area of scientific researches, for example,
statistical physics, physical chemistry
and theoretical biology. 
Simulations of lattice heteropolymers, 
which consist of several different types of monomers, 
are especially interesting because they are minimal models of protein folding \cite{SSK,DBYFYTC}.
Such simulations, however, 
often suffer from slow relaxations and metastability caused
by the competition between short-ranged interactions and connectivity
constraints among the monomers.
For other systems with metastability, such as a spin system which
exhibits the first-order phase transition,
the multicanonical ensemble method is known to work well.\cite{BN91}
But for the heteroprolymer simulations,
the multicanonical ensemble will not be a desired solution.
In fact,
even a self-avoiding walk, which is the simplest lattice polymer,
is not easy to generate \cite{Sokal95},
although no interaction energy is assumed between monomers other than the
constraint of the self-avoidingness.

In this letter, we propose a novel approach to the dynamical
Monte Carlo simulation of lattice polymers.
The present approach is a variant of Monte Carlo
algorithms based on extended ensembles
\cite{BN91,KT90,G91,MP92,Lee,I93,H96,BWA}
and can be applied to a wide range of models including lattice
heteropolymers and protein models on lattices.
Our starting point is the introduction of an artificial ensemble
that contains conformations with finite self-overlaps.  
With this relaxation of the self-avoidingness,
the conformations with self-overlaps play a role of ``bridges'' between
metastable states and
the rapid mixing of the Markov chain is expected.
In fact, Shakhnovich et al. have reported that the folding 
becomes considerably faster than usual
in the lattice protein model that allowed 
doubly and triply self-overlapping conformations.\cite{ADD}

But it is not easy to make adequate
portion of conformations self-avoiding. Consider a dynamical simulation of
self-avoiding walk. Once the self-overlap is allowed,
almost all conformations in the ensemble become non self-avoiding,
because self-overlapping conformations have 
larger entropy than self-avoiding ones.
Clearly we need penalties to self-overlap, whose prototype
is already seen in \cite{ADD}.
The problem is, however, how to choose a penalty term with proper
functional form and strength.
Our solution to this problem is similar to that in the multicanonical
ensemble. 
That is, the algorithm is 
designed to learn the appropriate values of the penalties 
for the violation of self-avoidingness in preliminary runs. 
After the values of the penalties are determined, 
a run for the measurement of physical quantities is performed.
By measuring physical quantities only for self-avoiding conformations, 
we get correct canonical averages.

The relaxation of self-avoidingness alone, however, is
not sufficient for an ensemble of good performance
when the models with attractive interactions are considered.
With these models, the polymer tends to collapse
at low temperatures, if finite self-overlaps are allowed.
Such behavior is evidently not desirable for our purpose, because
the severely collapsed conformations have smaller entropy and do not
work as good ``bridges''.
Thus we need an ensemble extended to {\it two directions }, 
that is,
the number of self-overlaps and the interaction energy.
There are several ways to realize such an ensemble.
Among them, we will discuss a particular type of ensemble ---
multi-self-overlap ensemble --- in detail. 

Let us consider a heteropolymer of length $N$ on a lattice $G$.
The energy of the polymer in a conformation $\Gamma$ is denoted
by $E(\Gamma)$.
Then the canonical distribution that we want to sample by Monte Carlo
simulations is
\begin{equation}
P_\beta(\Gamma) \propto \exp(-\beta E(\Gamma) ) {}_.
\end{equation}

In the conventional multicanonical algorithm \cite{BN91,UT}
we simulate the system with a modified probability
\begin{equation}
P_f(\Gamma) \propto \exp( - f(E(\Gamma)) ) {}_,
\end{equation}
where $f$ is a function only of $E$. 
Only the self-avoiding conformations are allowed as $\Gamma$.
The form of the function $f$
is tuned 
so that the marginal distribution of $E$ becomes 
as uniform as possible in a prescribed interval
$E_{min} < E < E_{max}$.
In actual implementations of the multicanonical algorithm, 
the appropriate values of $f(E)$ are learned by the iterative construction
of the energy histogram through preliminary runs. 
\cite{BN91,Lee}
Then the canonical average
at inverse temperature $\beta$ is calculated by the histogram reweighting.

In the multi-self-overlap ensemble, self-overlapping conformations
are allowed and
the probability is modified further as
\begin{equation}
P_{g}(\Gamma) \propto \exp(-g(E(\Gamma),V(\Gamma))).
\end{equation}
Here $V(\Gamma)$ is the effective energy associated with the 
self-overlaps in the
conformation $\Gamma$, which is defined as
\begin{equation}
V(\Gamma)= \sum_{i \in G'}  (n_i(\Gamma)-1)^2 {}_,
\end{equation}
where $n_i(\Gamma)$ is the number of the monomers on a
lattice point $i$, 
and $G'$ means a set of lattice points which are occupied by at least
one monomer.
When the polymer is in a self-avoiding
conformation, $V(\Gamma)= 0$, 
and when it is collapsed into
a pair of points, $V(\Gamma) \sim 2 \cdot (N/2-1)^2$;
values of $V(\Gamma)$ for any other conformations lie 
in between these two limiting values.
Here we assume that the definition of the original energy function $E(\Gamma)$ 
is extended to the conformations with self-overlaps
in a reasonable way.
We tune the values of $g(E,V)$ 
by a preliminary run
so that we get a sufficiently
flat bivariate marginal distribution of $(E,V)$ in a prescribed range
$E_{min} \leq E \leq E_{max}, \, 0 \leq V \leq V_{max}$.
Then a run for the measurement is performed.
In the numerical simulations in this paper, 
we use a method similar to the entropic sampling method.\cite{Lee} 
The reweighting formula for the present algorithm is 
\begin{equation}
  \langle A \rangle_\beta 
=
      \frac{ \sum_i'  A(\Gamma_i) P^{-1}_g(\Gamma_i) \exp(-\beta E(\Gamma_i) ) }
           { \sum_i'  P^{-1}_g(\Gamma_i) \exp(-\beta E(\Gamma_i) ) } {}_,
\end{equation}
where $\Gamma_i$ represents a conformation at $i$th Monte Carlo Step
and the summations are taken over the self-avoiding conformations.

Before discussing the numerical results,
we will touch on related works on extended ensembles. 
Urakami and Takasu investigated
lattice heteropolymers by means of the conventional
multicanonical algorithm \cite{UT}.
Bruce et al. \cite{BWA} calculated the free energy
difference between crystals of different lattice structures using
an extended ensemble. Although they treated very different systems, 
their idea had some common feature with ours. 

Now we consider a specific example as a touchstone of the algorighm,
that is,
the HP model \cite{LD89} of protein
on the square lattice.
This model consists of a self-avoiding polymer chain on a lattice
with two types of monomers (``amino acids'')
H(hydrophobic) and P(polar). 
A conformation of the polymer is indexed by
$ \mbox{\boldmath $r$} = \{r_i\} $, 
where $r_i$ is the vector of coordinates of
the $i$th monomer.  
The interaction $u$ between a pair of
monomers is defined as $u(H,H)=-1$, $u(H,P)=u(P,H)=u(P,P)=0$.
The energy $E( \mbox{\boldmath $r$}) $ 
of a conformation $\mbox{\boldmath $r$ } $
of a polymer
is written as
\begin{equation}
\label{energy}
E(\mbox{\boldmath $r$})=  \sum_{i<j}
u(S_i,S_j) \Delta(r_i-r_j) {}_,
\end{equation}
where $S_i \in \{H,P\}$ is the index of the type of $i$th monomer.
The factor $\Delta(r_i-r_j)$ takes the value 
1 
if and only if $r_i$ and $r_j$ are on adjoining lattice sites
but not consecutive along the chain.
We restrict the bond length between the consecutive monomers as 1,
even for self-overlap conformations.
Then,
we can simply use the same 
definition of the energy eq.~(\ref{energy}) for all the conformations.
We take $V_{max}$ = 4 throughout the following simulations.

Next, we select a set of ``elementary moves'' used in the simulations.
The following six moves are included,
which are widely used in dynamical Monte Carlo simulations
of lattice polymers:\cite{foot2} 
(1) one-bead flip, (2) $90^\circ$ end-bond rotation, 
(3) $180^\circ$ end-bond rotation, 
(4) $\pm90^\circ$ rotation, (5) $180^\circ$ rotation, 
(6) rotation and reflection.
We further add the following two new
moves which are possible only for self-overlapping conformations
to the move set:
(7a) ``jackknife'' shown in Fig.1a and (8a) ``loop overturn'' 
shown in Fig.1b.  
Note that the degrees of freedom for the moves of the conventional types
are increased when self-overlaps are allowed.

\begin{figure}%
\leavevmode
\epsfysize=5.5cm
\epsfbox{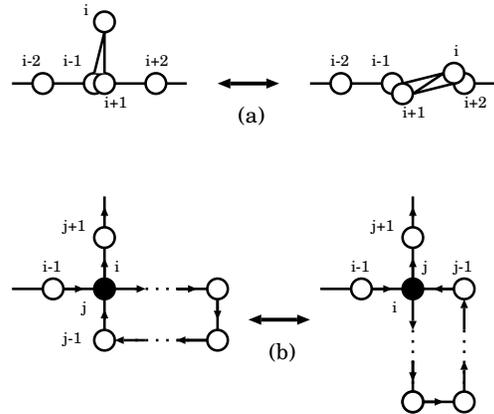}
\caption{New elementary moves:
       (a)''jackknife'':
         If the monomers $i-1$ and $i+1$ are on the same site,
         the monomer $i$ can be flipped into any site neighbouring to
         $i \pm 1$.
         In the right figure, the monomer $i$ is flipped to
         the site where the monomer $i+1$ is already situated. 
       (b)''loop overturn'':
          The monomers $i$ and $j$ are on the same site.
          If a loop is formed between the monomers $i$ and $j$
          which are on the same site,
          the loop can be overturned. }
\label{original_move}
\end{figure}

Let us turn to numerical results.
First we consider the
sequence $PH^2P^2H^2P^2H^3P^2HP$ of length 16,\cite{foot4} 
which is taken from the reference. \cite{SVMB96} 
This sequence has nine distinguishable ground states which are
not related with each other by symmetry.
We calculate the temperature dependence of the
``local field'' $\, f_i = \sum_{j}
u(S_i,S_j) \Delta(r_i-r_j)$ 
felt by the $i$th monomer. 
The results are shown in Fig.~2
along with the data calculated by the exact enumeration of 
all the conformations.
A good agreement
to the exact result supports the validity of the present algorithm.

\begin{figure}
\begin{center}
\leavevmode
\epsfysize=5.5cm
\epsfbox{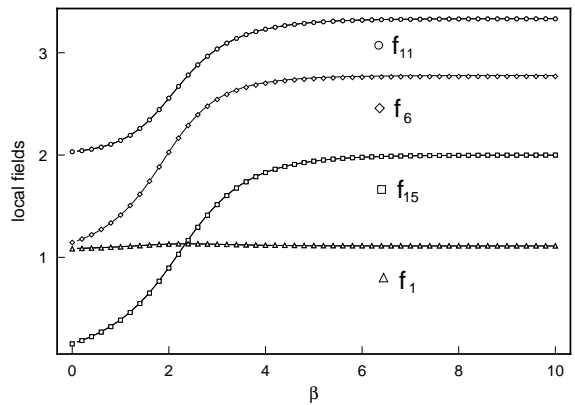}
\end{center}
\caption{Temperature dependance of the local fields for
          the heteropolymer of the length 16.
           Four typical local fields among 16 are shown.
           The symbols are the data calculated by the
            multi-self-overlap ensemble, and the line
            is the exact result calculated by the exact enumeration
            of all the conformations.}
\label{local fields}
\end{figure}

Next we consider the ``8-sticker'' sequence $P^3(HP^6)^7HP^3$ of 
length 56.\cite{UT}
The polymer with this sequence has a number of ground states. 
They can be classified into three large groups 
by the combinations of contacting H monomers \cite{foot3}.
Simulations of such polymer are evidently far from trivial. 
We compare the performance of the multi-self-overlap algorithm
with the conventional multicanonical algorithm
for this example.
Instead of the new elementary moves (7a) and (7b),
we use the following two moves 
in the conventional multicanonical algorithm:
(7b) $180^\circ$ crankshaft 
and (8b) three-bead J flip.

Time-series of the energy in both algorithms are plotted in Fig.3.
The definition of one Monte Carlo step is that
all types of the moves 
are tried once for all the monomers,
irrespective of whether the resulting conformation satisfies
the self-avoiding condition or not.
For the multi-self-overlap algorithm, 
the values of the energy only of self-avoiding conformations are shown.
It is clearly seen that inclusion of the self-overlapping conformation
accelerates the up-down itinerancy of energy very much.
In Fig.4, visit to each type of the ground states are recorded.
Jumps between the ground states of different types are evidently more frequent 
in the multi-self-overlap simulation than 
in the conventional multicanonical simulation. 
It suggests that the relaxation is much faster in the multi-self-overlap
algorithm.
We also make similar calculations by the multi-self-overlap algorithm
without using a new global move (8a).
As a result, we find that the performance slightly lowers than before 
but still is much higher than that of the conventional algorithm.

\begin{figure}
\begin{center}
\leavevmode
\epsfysize=7.5cm
\epsfbox{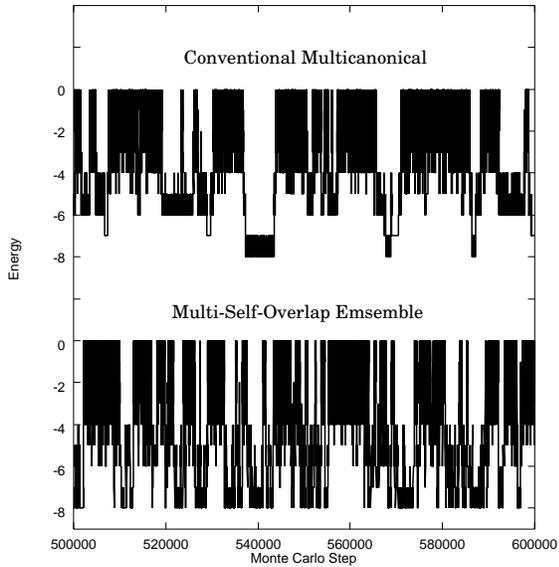}
\end{center}
\caption{Time series of the energy for the 8-sticker polymer
 of the length 56.
 The upper figure is for the conventional multicanonical ensemble,
 and the lower one is for the multi-self-overlap ensemble.
 Only the energy of the self-avoiding conformations are plotted
 in the lower figure; for counting the Monte Carlo steps,
 however, the self-overlapping conformations are also taken into
 account.}
\label{time_series_of_energy}
\end{figure}

\begin{figure}
\begin{center}
\leavevmode
\epsfysize=7.5cm
\epsfbox{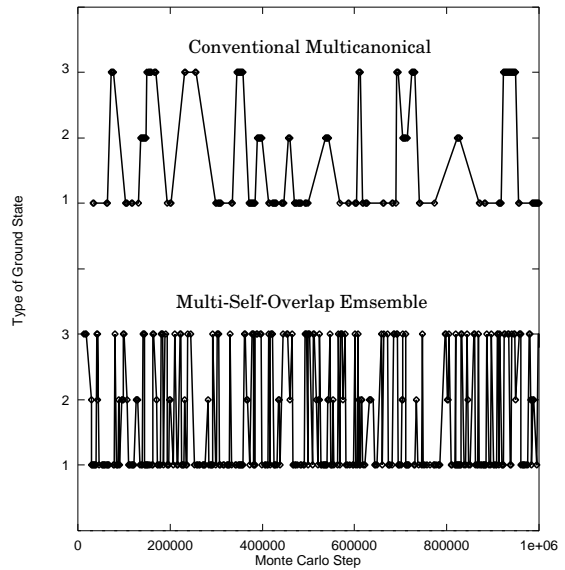}
\end{center}
\caption{Itinerancy among the ground states of three types against
the Monte Carlo steps.
The upper figure is for the conventional multicanonical ensemble,
and the lower one is for the multi-self-overlap ensemble.
The ordinate indicates the type of the ground states.}
\label{Ground_State}
\end{figure}

So far, we discuss the performance in the measurement runs.
On the other hand, our experience shows
that the multi-self-overlap algorithm has better performance even 
in the preliminary runs,
i.e., the tuning of the weight requires much shorter CPU time
than in the conventional multicanonical algorithm, 
despite the two-dimensional nature of the histogram to be constructed 
in the preliminary runs.
Actually, we need a few iterations
each of which consists only of 50000 Monte Carlo steps 
to obtain sufficiantly uniform histogram
by the multi-self-overlap algorithm.
On the other hand, 
at least $10^6$ Monte Carlo steps are needed in each iteration
by the conventional multicanonical algorithm.

Finally, we briefly discuss a different implementation
of our idea.
Here we construct an extended ensemble
following a method similar to the exchange Monte Carlo algorithm
\cite{H96,I93}
(equivalently, Metropolis-coupled Markov chain Monte Carlo algorithm
\cite{G91}, 
or, Time-homogeneous annealing algorithm \cite{KT90}) 
.
Instead of a series of canonical ensembles
with different temperatures in the usual exchange algorithm \cite{H96},
we introduce a series of $n$ modified distributions
$\{P_k(\Gamma_k) \}\,
(1 \leq k \leq n )$, where
\begin{equation}
P_k(\Gamma_k) 
\propto \exp( - \beta_k E(\Gamma_k) - \lambda_k V(\Gamma_k)  \, ) 
\end{equation}
with different degrees of penalties $\{\lambda_k\}$ to self-overlap
and inverse temperatures $\{ \beta_k \}$. 
The value of $\lambda_n$ is set large enough to prevent 
self-overlap and 
$\beta_n$ as the inverse temperature $\beta$ originally 
required in the problem. 
The exchange algorithm is defined as follows:
$n$ dynamical Monte Carlo simulations are performed
in parallel, each for $k$th parameter set
$(\lambda_k, \beta_k)$,
with a pair of the conformations
$\Gamma_k$ and $\Gamma_{k'}$ being exchanged at a
prescribed interval according to a probability
\begin{equation}
r = \max{ \left \{ 1,  \frac{P_k(\Gamma_{k'}) \cdot P_{k'}(\Gamma_{k})}
{P_k(\Gamma_{k}) \cdot P_{k'}(\Gamma_{k'})}  \right \} } .
\end{equation}
The equilibrium distribution of this coupled Markov chain
is the simultaneous distribution
$\Pi_kP_k$.
Then, a sample from the distribution $P_n$
is regarded as a sample from the desired distribution. 

In this paper, we proposed a novel family of dynamical Monte Carlo 
algorithms for the simulation of lattice polymers. 
The essence of our algorithms is the introduction of extended ensembles
in which the self-avoiding conditions are systematically weakened.
We discussed two different implementations of the idea. 
Both ensembles contains an adequate portion of 
higher energy conformations as well as self-overlapping conformations. 
We perform the numerical simulations using one of them, 
the multi-self-overlap ensemble algorithm.
It achieved superior performance compared with
the conventional multicanonical algorithm 
in the hard problem of  8-sticker HP polymers. 
The following point should be stressed:
Although
only a portion of the generated conformations satisfy the self-avoiding
condition (about 1/5 in the simulations presented above, 
since $V_{max}=4$),
we still can get much more statistically independent samples
compared with the conventional multicanonical ensemble,
because the relaxation is accerelated very much.
We also note that the proposed algorithms {\it correctly } 
reproduce the canonical averages at finite temperature
when the averages are taken over self-avoiding conformations.

We believe that the present algorithm is powerful and general tool
to investigate lattice heteropolymers.
An interesting application will be
in the design of lattice proteins \cite{SVMB96}. 
Research in this direction is now in progress and will be published 
in the forthcoming paper \cite{IKT98}.
Finally, we point out that the idea behind the multi-self-overlap ensemble
is easily extended to simulations of off-lattice polymer models.
Application to the finite temperature simulations of the realistic 
protein models
will also be an interesting problem, where
the conventional multicanonical ensemble method is currently used.
\cite{HO}

\acknowledgements
We would like to thank Y.~Akutsu, Y.~Okabe, T.~Kawakatsu, M.~Takasu and
K.~Tokita for
fruitful discussions and comments.
The work is supported in part by Grant-in-Aid for Scientific Research on 
Priority Areas from The Ministry of Education, Science, Sports and Culture.


\begin{thebibliography}{99}
\bibitem{foot2}
Description of the conventional moves are found in ref.\cite{Sokal95}.
\bibitem{foot4}
We use a convention to contract the notation.
For example, $H^3 P^2$ means $HHHPP$ and $(HP)^2$ means
$HPHP$.
\bibitem{foot3} 
Each ground state conformation has two H cores,
each of which consists of four H monomers making $2\times 2$ square.  
Type 1 ground states
have H cores of (4,11,18,25) monomers and (32,39,46,53) monomers.
Type 2 consists of two types of H core structures:
(4,11,18,53) and (18,25,32,39) or (4,39,46,53) 
and (11,18,25,32). 
In type 3 ground states, H cores are (4,11,46,53) and (18,25,32,39).
\end{thebibliography}
\end{document}